\documentclass[preprint,12pt]{elsarticle}

\usepackage{amssymb}

\usepackage{amsmath}

\newcommand{\beq}{\begin{equation}}
\newcommand{\eeq}{\end{equation}}
\newcommand{\bea}{\begin{align}}
\newcommand{\eea}{\end{align}}

\newcommand{\var}{\mathrm{Var}}

\newcommand{\ev}{\mathrm{E}}

\usepackage[usenames,dvipsnames]{color}

\newcommand{\snp}[1]{{\textbf{\color{red} #1}}}
\newcommand{\evrel}{\chi}

\journal{Theoretical Population Biology}

\begin{document}

\begin{frontmatter}



\title{
A generalized Watterson estimator for next-generation sequencing: 
from trios to autopolyploids
}

\author[upmc,cdf]{Luca Ferretti\corref{cor1}}
\ead{luca.ferretti@gmail.com}
\author[crag]{Seb\'astian E. Ramos-Onsins}                                   
\cortext[cor1]{Corresponding author}
\address[upmc]{Syst\'ematique, Adaptation et Evolution (UMR 7138), UPMC Univ Paris 06, CNRS, MNHN, IRD, Paris, France}
\address[cdf]{CIRB, Coll\`ege de France, Paris, France}
\address[crag]{Centre for Research in Agricultural Genomics (CRAG) CSIC-IRTA-UAB-UB, Edifici CRAG, Campus Universitat Aut\`onoma.  Bellaterra 08193, Spain}

\begin{abstract}
Several variations of the Watterson estimator of variability for Next Generation Sequencing (NGS) data have been proposed in the literature. 
We present a unified framework for generalized Watterson estimators based on Maximum Composite Likelihood, which encompasses most of the existing estimators. 
We propose this class of unbiased estimators as generalized Watterson estimators for a large class of NGS data, including pools and trios. 
We also discuss the relation with the estimators that have been proposed in the literature and show that they admit two equivalent but seemingly different forms, 
deriving a set of combinatorial identities as a byproduct. 
Finally, we give a detailed treatment of Watterson estimators for single or multiple autopolyploid individuals.
\end{abstract}


\end{frontmatter}



\section{Introduction}
\label{intro}

The rescaled mutation rate per base $\theta$ plays an important role in 
population genetics models. Its definition is $\theta=2pN_e\mu$ where $\mu$ denotes the mutation rate per base, $N_e$ the effective population size and $p$ the ploidy of the species. In the context of the Standard Neutral Model (SNM) at low mutation rate, 
commonly used estimator include the Watterson estimator \cite{watterson1975number}, based on the number of segregating sites $S$ in a sample from the population, and the pairwise nucleotide diversity $\Pi$ \cite{tajima1983evolutionary}, defined as the average number of differences per base between two individual sequences. Despite its simple interpretation and its popularity, $\Pi$ is an inconsistent estimator of $\theta$, while the Watterson estimator is a good estimator since it is unbiased and it corresponds to the Maximum Composite Likelihood estimator for  $\theta$. Furthermore, as shown in \cite{roychoudhury2010sufficiency}, it is a sufficient statistics for $\theta$ for large sequences. 
For unequal mutation rates between different alleles, the Watterson estimator actually measures the net rate of mutation towards different alleles, rescaled by population size.

Today,  variability analyses on a genome-wide scale can be easily done by Next Generation Sequencing (NGS) technologies. NGS technologies can sequence a single complete genome at relatively high redundancy, but the sequencing of many samples increases substantially the cost of the experiment. Several strategies have been used to obtain sequence data from many individuals, from restriction reduced libraries to pooled samples. These strategies may reduce substantially the coverage across the genome (i.e. how many regions are sequenced) and the sequence redundancy per nucleotide base (i.e. how many sequences cover each position). From this point of view, the study of variability on NGS data can be seen as a missing data problem, 
where information about several samples is missing at covered positions. The Watterson $\theta$ estimator has been generalized to missing data by Ferretti et al. \cite{ferretti2012neutrality}, but it covers just a limited number of cases.


Estimators of variability based on NGS data should take into account the relatively high probability of sequencing errors in order not to overestimate the number of variants. Sequence redundancy at a given position is fundamental for detecting and removing errors but also for detecting correctly homozygote or heterozygote positions in  diploid individuals. 
NGS estimators of variability were first proposed by Lynch \cite{lynch2008estimation} for a single diploid individual using the sequence error rate and the number of reads observed at each variant. 
Hellmann et al. \cite{hellmann2008population} and Jiang et al. \cite{jiang} proposed Watterson estimators for multiple individuals taking into account the combinatorics of reads from different individuals and homologous chromosomes. 
Later, Futschik and Schl\"otterer \cite{Futschik:2010fk}  and Ferretti et al. \cite{ferretti2013} developed variability estimators for pooled sample data using Method of Moments (MM) and Maximum Composite Likelihood (MCL) methods, respectively. 
We observe that  for the  analysis of a single diploid individual, they all reduce to the same estimator except for \cite{Futschik:2010fk} (the differences between this estimator and the MCL have been detailed in \cite{ferretti2013}). In this work, we study the relations between these estimators and present a common framework for estimators of variability. We develop new Watterson estimators for other kinds of NGS data, like trios or autopolyploids.


In our framework, we deal with all these cases as special instances of a general, unified approach. Data are represented by NGS or Sanger sequences coming from several units. 
Units can be haploid, diploid or polyploid individuals or pools, each one containing a different number of lineages.  
In section \ref{sec_genwatt}, we derive Maximum Composite Likelihood estimators for a generic class of data. These MCL estimators are not unbiased, but we show that they can be well approximated  by unbiased estimators that share the same functional form as the Watterson estimator. In section \ref{sec_ngs}, we propose these unbiased estimators as generalized Watterson estimators for a large class of NGS data, such as multiple haploid/diploid/polyploid individuals, pools, trios, inbred lines, and combinations of them. We present explicit formulae for these estimators in section \ref{sec_cases}.
We discuss their relation with other estimators that have been proposed in the 
literature, showing in section \ref{sec_dual} that many of our estimators admit two equivalent but seemingly different forms. As a byproduct, this equivalence implies a set of combinatorial identities. 
Finally, in section \ref{sec_poly} we treat in details the case of single and multiple autopolyploid individuals and we provide the Watterson estimator for autopolyploids.

\section{General Watterson estimators} \label{sec_genwatt}

\subsection{Maximum Composite Likelihood estimators} \label{sec_mcle}


Composite Likelihood Estimation of parameters has been extensively used in population and quantitative genetics for estimating the linkage disequilibrium among positions \cite{Devlin:1996vn,Weir:1979kx} and for estimating evolutionary parameters as the level of variation \cite{hellmann2008population}, the recombination rate \cite{hudson2001two}, \cite{McVean:2002}, 
the strength of positive selection \cite{KimStephan2002, ZhuBustamante2005} or demographic parameters \cite{Garrigan:2009}. 
Maximum Composite Likelihood is an appropriate method to estimate the nucleotide variability across large regions because it has minimum mean squared error for large recombining regions, since in this case the Composite Likelihood is a good approximation for the exact likelihood and therefore the estimator is approximately asymptotically efficient. 
Furthermore, it turns out to be based on the same statistics as the Watterson estimator (the total number of segregating sites $S$) and 
it actually reduces to the Watterson estimator if the data are represented by complete sequences. 

In this paper, we consider allelic variants represented by segregating sites, or Single Nucleotide Polymorphisms (SNPs), but the methods can be applied to generic variants with low mutation rate.

For each site, there are features that depend on an eventual SNP (for example, allele frequencies) and features that do not depend on the allelic content (for example the read depth, i.e. the number of sequences that contain data for a given site). We summarize the SNP features with the index $\xi\in \Xi$ and the features of each site that do not depend on the allelic content with  the index $\varphi\in \Phi$. Both indices indicate 
mutually exclusive, collectively exhaustive features.

We denote by $p_{\varphi,\xi}(\theta)$ the probability that a site with features $\varphi$ contains an observed SNP with features $\xi$ for the sample studied. 
For small values of $\theta$ 
(that is, in the infinite site model)
, we can expand it in Taylor series and using the fact that there are no SNPs without mutations, i.e. $p_{\varphi,\xi}(0)=0$, we find that these probabilities are proportional to $\theta$ multiplied by a quantity that depends on the population model and the sequencing setup:
\beq
p_{\varphi,\xi}(\theta)\simeq\theta Z_{\varphi,\xi}\ .\label{eq_p}
\eeq
We denote by $S_{\varphi,\xi}$ the number of segregating sites with features $\xi$ in positions with features $\varphi$, and by $L_\varphi$ the number of sites with features  $\varphi$. We also define the quantities $S_\varphi=\sum_{\xi\in \Xi}S_{\varphi,\xi}$ and $Z_\varphi=\sum_{\xi\in \Xi}Z_{\varphi,\xi}$, 
the total number of segregating sites $S=\sum_{\varphi\in \Phi}S_\varphi$ and total length $L=\sum_{\varphi\in \Phi}L_\varphi$.

Under the composite approximation, all sites are independent. The Composite Likelihood is therefore the simple product of probabilities:
\beq
CL(\theta)=\left[ \prod_{\varphi\in \Phi} \prod_{\xi\in \Xi}  p_{\varphi,\xi}(\theta)^{S_{\varphi,\xi}} \right]\cdot \left[\prod_{\varphi\in \Phi} \left( 1-\sum_{\xi\in \Xi} p_{\varphi,\xi}(\theta)\right)^{L_\varphi-S_\varphi} \right] \ . \label{cl}
\eeq
Substituting eq. (\ref{eq_p}) for $p_{\varphi,\xi}(\theta)$  and taking the log, we obtain 
\beq
\log(CL(\theta))=S\log(\theta)+\sum_{\varphi\in \Phi}\sum_{\xi\in \Xi}S_{\varphi,\xi}\log(Z_{\varphi,\xi})+\sum_{\varphi\in \Phi}(L_\varphi-S_\varphi)\log(1-\theta Z_\varphi)\ .\label{logcl}
\eeq 
For $S>0$, the loglikelihood is always negative and tends to $-\infty$ both for $\theta\rightarrow 0$ and $\theta\rightarrow 1/\max_{\varphi\in \Phi}(Z_\varphi)$, so the maximum can be obtained from the zeros of the first derivative of the $\log(CL)$ in equation (\ref{logcl}). After rearrangements, we obtain the equation that defines the Maximum Composite Likelihood Estimator (MCLE):
\beq
L=\sum_{\varphi\in \Phi}\frac{L_\varphi-S_\varphi}{1-\hat{\theta}_{MCLE} Z_\varphi}\ ,\label{mcle}
\eeq
which is valid for all values of $\theta< 1/\max_{\varphi\in \Phi}(Z_\varphi)$ since the second derivative in $\theta=\hat{\theta}_{MCLE}$ is negative if $\sum_{\varphi\in \Phi}(L_\varphi-S_\varphi)Z_\varphi/(1-\hat{\theta}_{MCLE} Z_\varphi)^2>0$, which is always verified.

For the simplest case of the original Watterson estimator \cite{watterson1975number}, all sites are equivalent and there is no site feature $\varphi$, so the above MCLE equation (\ref{mcle}) can be easily rewritten as $L=(L-S)/(1-\hat{\theta}_{MCLE}Z)$, i.e. $\hat{\theta}_{MCLE}=\hat{\theta}_{W}={S}/({LZ})$. The SNP features $\Xi$ correspond simply to the derived allele counts $i=1\ldots n-1$ and the probability of a SNP of frequency $i/n$ is related to the expected frequency spectrum $\xi_i$ - defined as the count of SNPs of frequency $i/n$ in the sample - by $p_i(\theta)=\ev(\xi_i)/L$. For the standard neutral model, $p_i(\theta)=\theta/i$, therefore $Z=\sum_{i=1}^{n-1}Z_i=\sum_{i=1}^{n-1}p_i(\theta)/\theta$ is given by the harmonic number $a_n=\sum_{i=1}^{n-1}1/i$. Then the MCLE in this case corresponds precisely to the unbiased Watterson estimator $\hat{\theta}_W=S/(L\sum_{i=1}^{n-1}1/i)=S/(La_n)$.

The estimator (\ref{mcle}) is defined implicitly, so it is not easy to use. An explicit, approximate MCL estimator can be derived in two equivalent ways: either (i) by expanding equation (\ref{mcle}) at first order in the small  parameters $\theta$ and $S_\varphi/L_\varphi$ with constant ratio $S_\varphi/(\theta L_\varphi$), or (ii) by taking the small $\theta$, large $L$ limit of the likelihood (\ref{cl}) with $\theta L$ and $L_\varphi/L$ constant;  
in this limit, the $S_{\varphi,\xi}$ are Poisson distributed random variables with mean $ L_\varphi\theta Z_{\varphi,\xi}$ 
\beq
CL(\theta)\simeq\prod_{\varphi\in \Phi} \prod_{\xi\in \Xi}  \frac{(L_\varphi \theta Z_{\varphi,\xi})^{S_{\varphi,\xi}}}{S_{\varphi,\xi}!}e^{-L_\varphi \theta Z_{\varphi,\xi}}=\theta^Se^{-\theta\sum_{\varphi\in \Phi} L_\varphi Z_\varphi}\left[\prod_{\varphi\in \Phi} \prod_{\xi\in \Xi}\frac{(L_\varphi Z_{\varphi,\xi})^{S_{\varphi,\xi}}}{S_{\varphi,\xi}!}\right]
\eeq
and since the dependence on $\theta$ lies in the first term which is a function of the statistics $S$ only,  $S$ is a sufficient statistics for $\theta$ by the Fisher-Neyman factorization theorem, as already observed in \cite{roychoudhury2010sufficiency}. 

Both ways lead to the same estimator. We define the resulting approximate MCL estimator as the generalized Watterson estimator:
\beq
\hat{\theta}_{W}=\frac{S}{\sum_{\varphi\in \Phi}L_\varphi Z_\varphi}\ .\label{watt}
\eeq
This estimator depends only on the total number of segregating sites, like the original Watterson estimator, since $S$ is a sufficient statistics for small $\theta$. Furthermore, it is unbiased. In fact, $\ev(S)=\sum_{\varphi\in \Phi}\sum_{\xi\in \Xi}\ev(S_{\varphi,\xi})=\sum_{\varphi\in \Phi}\sum_{\xi\in \Xi}L_\varphi p_{\varphi,\xi}(\theta)=\sum_{\varphi\in \Phi}\theta L_\varphi Z_\varphi$. 

Both the implicit estimator in equation (\ref{mcle}) and the formula (\ref{watt}), which is unbiased, can be used. The relative error of the Watterson estimator (\ref{watt}) with respect to the MCLE (\ref{mcle}) is very small.  By expanding eq. (\ref{mcle}) at first order in $\hat{\theta}_{MCLE}-\hat{\theta}_W$ and at first nonzero order in $\hat{\theta}_W$ and $S_\phi/(L_\phi Z_\phi)$, we obtain an error estimate of order $\theta$ multiplied by a weighted covariance between $Z$ and the relative fluctuations of $\hat{\theta}_{W}$ for different $\varphi$s:
\beq
\frac{\hat{\theta}_{MCLE}-\hat{\theta}_{W}}{\hat{\theta}_{W}}\simeq -\hat{\theta}_{W} \sum_{\varphi\in \Phi}\frac{L_\varphi Z_\varphi}{\sum_{\varphi'\in \Phi}L_{\varphi'}Z_{\varphi'}} \left(Z_\varphi-\frac{\sum_{\varphi'\in \Phi}L_{\varphi'}Z_{\varphi'}}{L}\right)\left(\frac{S_\varphi/(L_{\varphi}Z_{\varphi})-\hat{\theta}_{W}}{\hat{\theta}_{W}}\right)
\eeq
up to terms of order $(\hat{\theta}_W,S_\phi/(L_\phi Z_\phi))^2$. 
This error is usually negligible, since 
the r.h.s. is suppressed by a factor of $\theta\ll 1$; furthermore it has mean 0, since it correlates the fluctuations of the mutation process with the fluctuations of the sequencing process, but the two processes are independent. 

The only information needed to compute (\ref{mcle}) or (\ref{watt}) are the factors $Z_{\varphi}=\sum_{\xi\in \Xi}p_{\varphi,\xi}/\theta$, which depends both on the model and on the sequencing setup. In the rest of the paper, we will specialize these factors for several combinations of NGS data.


\section{Application to NGS and sequence data} \label{sec_ngs}

\subsection{The data: sequences aligned to a reference genome}

In this section we deal with combinations of complete sequences, genotypes and NGS data from different sources in a unified way. Our data are represented by reads, sequences or genotypes\footnote{Some genotyping methods (like DNA microarrays) preselect the possible SNPs or the positions containing a segregating site. These methods give biased estimates of variability and cannot be meaningfully combined  with unbiased methods like Sanger or NGS.}, aligned to a reference genome. Each read/sequence/genotype originates from a single unit: units can be individuals of different ploidy, or pools of individuals. Complete sequences are considered as sequences coming from an haploid unit, 
so the two complete sequences of the two homologous chromosomes from a diploid organism are equivalent to two different haploid units.

We denote by $U$ the number of units. The features associated to the units are (i) the number of copies of homologous chromosomes present in each unit and (ii) the evolutionary relationships between the units. We denote the number of copies of homologous chromosomes in the $i$th unit by $c_i$, $i=1\ldots U$. 
Diploid individuals will have $c_i=2$, polyploid individuals will have $c_i$ equal to their ploidy and pools will have $c_i$ equal to the number of individuals in the pool multiplied by their ploidy. We denote the set of numbers $\{c_i\}_{i=1\ldots U}$ by $\{c\}$. 

The evolutionary relationships (denoted here by the generic symbol $\evrel$) include all the available information relevant for the probabilities that some of the sequenced chromosome derived by the same lineage, either because they are actually from the same individual or because are identical by descent. For example, two different pools could contain two genetically identical individuals, or two individuals could be parent and offspring, or a single diploid individual could originate from a single inbred line with given inbreeding coefficient, and so on. 

We assume that the number of NGS reads covering each position depends only on the sequencing process and not on the allelic composition of the sequence. 
In this case, we associate to  each position $x$ the read depth of the $i$th unit  $r_i(x)$, $i=1\ldots U$, i.e. the number of reads or sequences from the $i$th unit that cover position $x$. For Sanger sequences and genotyping, we define an ``effective read depth' $r_i(x)=0$ for positions with missing data and $r_i(x)=+\infty$ otherwise. We denote the set of read depths $\{r_i\}_{i=1\ldots U}$ by $\{r\}$. 

An example of the data is given in Table \ref{tab1}.

\begin{table}
\begin{center}
\begin{tabular}{ l r | c c c c c c c c c c c c }
Reference &   & A&C&A&C&G&T&A&A&T&C&G&C \\ \hline
Sanger:     & (unit 1)  & A&C&A&C&\snp{G}&T&T&A&\snp{A}&C&G&\snp{C} \\
                   & (unit 2) & A&C&A&C&\snp{C}&T&-&A&\snp{T}&C&G&\snp{C} \\
                    & &&&&&&&&&&&& \\
Genotyping:  & (unit 3) & $\frac{\mathrm{A}}{\mathrm{A}}$&$\frac{\mathrm{C}}{\mathrm{C}}$& -
&$\frac{\mathrm{C}}{\mathrm{C}}$&\snp{$\frac{\mathrm{C}}{\mathrm{G}}$}&$\frac{\mathrm{T}}{\mathrm{T}}$&$\frac{\mathrm{T}}{\mathrm{T}}$&$\frac{\mathrm{A}}{\mathrm{A}}$&\snp{$\frac{\mathrm{T}}{\mathrm{T}}$}&$\frac{\mathrm{C}}{\mathrm{C}}$&$\frac{\mathrm{G}}{\mathrm{G}}$&\snp{$\frac{\mathrm{C}}{\mathrm{G}}$} \\
                    &  &&&&&&&&&&&& \\
NGS reads: & (unit 4) & &&A&C&\snp{G}&T&T&A&&&& \\
                    & (unit 4) & A&C&A&C&\snp{C}&&&&&&& \\
                    & (unit 4) & &&&C&\snp{C}&T&T&A&\snp{A}&&& \\
                    & (unit 5) & &&&&&&T&A&\snp{A}&C&G&\snp{C} \\
                    & &&&&&&&&&&&& \\ \hline
Read depths:          &         &&&&&&&&&&&& \\
unit 1: $c_1=1$  &$r_1=$& $\infty$& $\infty$& $\infty$& $\infty$& $\infty$ & $\infty$& $\infty$& $\infty$& $\infty$& $\infty$& $\infty$& $\infty$ \\
 unit 2: $c_2=1$ & $r_2=$ & $\infty$& $\infty$& $\infty$& $\infty$& $\infty$ & $\infty$& $0$& $\infty$& $\infty$& $\infty$& $\infty$& $\infty$ \\
  unit 3: $c_3=2$ & $r_3=$ & $\infty$& $\infty$& $0$& $\infty$& $\infty$ & $\infty$& $\infty$& $\infty$& $\infty$& $\infty$& $\infty$& $\infty$ \\
 unit 4: $c_4=2$ &$r_4=$& 1&1&2&3&3&2&2&2&1&0&0&0 \\
 unit 5: $c_5=2$ &$r_5=$& 0&0&0&0&0&0&1&1&1&1&1&1 \\      
\end{tabular}
\end{center}
\caption{Example (not real) of data from five sequencing units from different sources and technologies. The data come from four diploid individuals, of which two are parent and child. The data are aligned to the reference genome (the sequence at the top). There are two complete Sanger sequences (units 1 and 2) with a missing base in the 7th position, one sequence of genotypes (unit 3) with a missing genotype at the 3rd position, and four NGS read (three coming from unit 4 and one from unit 5). The allelic content of the positions with SNPs is shown in bold red. At the bottom, we report the ``effective' read depths for all units. All units are diploid individuals (therefore $c=2$). Units 1 and 2 are the two homologous sequences of the parent (hence $c_1=c_2=1$) and unit 5 comes from the child ($c_5=2$), while units  3 and 4 come from unrelated individuals ($c_3=c_4=2$).}\label{tab1}
\end{table}

\subsection{Estimators for Next Generation Sequencing}

We consider NGS data like the ones described above.  We can derive the general Watterson estimator for this case by using the definition (\ref{watt}) with $\varphi=\{r\}=\{r_i\}_{i=1\ldots U}$, the set of read depths of the different units at a given site. We use the short form $\{r\}=\{r_i\}_{i=1\ldots U}$ and $\{c\}=\{c_i\}_{i=1\ldots U}$ for the information about the site features. $Z$ can be computed by
conditioning on the number of unrelated homologous chromosomes or lineages which actually contribute to the data, 
denoted by $j$, and then averaging over $j$:
\beq
\sum_{\xi\in\Xi}p_{\{r\},\xi}=\theta Z_{\{r\}}=
\sum_{j=2}^{\infty}P(\mathrm{SNP}|\evrel,\{c\},\{r\},j)\cdot P_c(j|\evrel,\{c\},\{r\})
\eeq
where $P(\mathrm{SNP}|\ldots)$ is the probability of observing a SNP among $j$ independent lineages and $P_c(j|\ldots)$ is the distribution of $j$,
which 
depends also on $\{c\}$ and on their relationships $\evrel$.
The probability of observing a SNP among $j$ independent lineages depends just on $j$, not on the redundancy of each lineage in the sequences, and on the expected site frequency spectrum $\ev(\xi_k|j)$ of the model, being equal to $P(\mathrm{SNP}|j)=\sum_{k=1}^{j-1}\ev(\xi_k|j)/L$. In the case of the standard neutral Wright-Fisher model we have $\ev(\xi_k|j)=\theta L/k$ and therefore $P(\mathrm{SNP}|j)=\theta\sum_{k=1}^{j-1}1/k=\theta a_j$. 

The general Watterson estimator for the Standard Neutral Model (SNM) is then
\beq
\hat{\theta}_W=\frac{S}{\sum_{\{r\}}L_{\{r\}}\sum_{j=2}^{\infty} P_c(j|\evrel,\{c\},\{r\})\cdot a_j}\ .\label{thetaw}
\eeq

This form was found by \cite{hellmann2008population} and \cite{perez2009,ferretti2013} in specific cases, but it holds for a very large class of data as shown. In the next section we will find the expression of $P_c(j|\ldots)$ for the most common sequencing setups. Note that this estimator does not take into account sequencing errors. The treatment of sequencing error will be presented in the Discussion.

Complete sequences and genotyping data can be also analyzed by the above formulae, provided that an effective read depth $r=+\infty$ is considered for positions with data, and $r=0$ for positions with missing data. In fact, if there are missing data, the information is equivalent to the absence of NGS data ($r=0$), while presence of data means that the genotype is known with certainty, as it would be with large read depth for NGS data ($r=\infty$).

Finally, the general Watterson estimator for an arbitrary model with spectrum $\ev(\xi_k|j)=\theta L \bar{\xi}_{k,j}$  is
\beq
\hat{\theta}_W=\frac{S}{\sum_{\{r\}}L_{\{r\}}\sum_{j=2}^{\infty} P_c(j|\evrel,\{c\},\{r\})\sum_{i=1}^{j-1}\bar{\xi}_{i,j}}\ .\label{thetawgen}
\eeq

\subsection{A simple example: data from a single diploid individual}

In this section we present the case of a single diploid individual [Lynch]. The single individual represents a single unit $U=1$, and being diploid (assuming unrelated parents) we have two homologous sequences, therefore $c_1=2$. Assuming that each read is extracted at random from one of the two homologous sequences, the estimator is specified by the probabilities
\begin{align}
&P_c(j=1|c_1=2,r_1)=2^{-r_1+1}\\
&P_c(j=2|c_1=2,r_1)=1-2^{-r_1+1}
\end{align}
since the probability of extracting all $r_1$ reads from a given sequence is $2^{-r_1}$. Therefore the estimator from equation (\ref{thetaw}), taking into account the relevant harmonic factors  $a_1=0$ and $a_2=1$, is
\beq
\hat{\theta}_W=\frac{S}{\sum_{r_1=2}^\infty L_{r_1}(1-2^{-r_1+1})} \ .
\eeq

As an example, consider the sequences in Table \ref{tab_1d}. In the sequence of length $L=7$ there are $S=2$ segregating sites. There are bases with read depth 1, 2 and 3 and their numbers are $L_1=1$, $L_2=4$ and $L_3=2$ respectively. The value of the Watterson estimator for this sequence is therefore $\hat{\theta}_W=2/(1\cdot 0+4\cdot 1/2+ 2\cdot 3/4)=4/7\simeq 0.57$.

\begin{table}
\begin{center}
\begin{tabular}{ l | c c c c c c c }
Reference &   A&G&A&C&C&A&T \\ \hline
NGS reads: &   &&\snp{A}&C&C&\snp{T}&T \\
&   &G&\snp{C}&C&C&\snp{A}&T \\
&   A&G&\snp{C}&C& & &  \\                    \hline
Read depths: $r_1=$         & 1 &2 &3 &3 &2 &2 &2  
\end{tabular}
\end{center}
\caption{Example  (not real) of data from a single diploid individual.  The reads are aligned to the reference genome (the sequence at the top). Positions with SNPs are shown in bold red. }\label{tab_1d}
\end{table}

\section{Distributions of the number of sequenced lineages}\label{sec_cases}
As discussed in the previous sections, the distribution of the number of sequenced lineages $P_c(j|\ldots)$ is actually enough to define the Watterson estimator. Before deriving its expression for a number of cases, we introduce some notation.

We denote the Stirling numbers of second kind for $j$ sets from $r$ objects by $S(r,j)$. We define the probability distribution
\beq
P^*(j|c,r)=\frac{c!\ S(r,j)}{(c-j)!\ c^{r}}
\label{probps}
\eeq
that corresponds to the probability of extracting exactly $j$ different objects with $r$ extractions (with repetitions) from a set of $c$ objects \cite{ferretti2013}. In fact, the number of possible extractions from a set of $c$ objects is $c^r$, while the number of extractions of precisely $j$ objects is given by the product of the number of ordered choices of $j$ objects out of $c$, that is $c!/(c-j)!$, multiplied by the number of ways to distribute the $j$ objects across $r$ extractions $S(r,j)$. Since all extractions are equiprobable, the ratio gives the the probability (\ref{probps}).  This equation will often appear in the formulae for the distribution of the number of lineages.

We  denote by $I(x)$ the indicator function that takes the value $1$ if $x$ is true and $0$ otherwise. We also denote by $\delta_{i,j}$ the Kronecker delta, that is, the identity matrix $\delta_{i,j}=I(i=j)$. Note that $P^*(j|c,0)=\delta_{j,0}$.

\subsection{General case: independent lineages}\label{probindlin}

Assume that all lineages in these units are independent. This corresponds to sequencing many unrelated individuals in a population without inbreeding. 
If there are $U$ units, $c_i$ is the number of lineages/homologous chromosomes in the $i$th unit, and $r_i$ is the number of reads/sequences coming from the $i$th unit, the probability $P_c(j|\{c\},\{r\})$ in the Watterson estimator is
\beq 
P_c(j|\{c\},\{r\})=\sum_{i_1=0}^{c_1}\ldots\sum_{i_{U}=0}^{c_{U}}I\left(j=\sum_{p=1}^{U} i_p\right)\prod_{q=1}^{U} P^*(i_q|c_q,r_q)\label{indlineages}
\eeq
which is a product of probabilities for each unit of sequencing $i_1\ldots i_U$ chromosomes respectively, summed over all combinations resulting in $j$ independent chromosomes.
 
In Section \ref{sec_dual} we will present an alternative form for these Watterson estimators. In the rest of this section we specialize the expression (\ref{indlineages}) to the most common scenarios.

\subsubsection{Multiple haploid individuals}

In this case all individuals have $c_i=1$, therefore $P^*(i_q|c_q=1,r_q)=I(i_q=I(r_q>0))$ and the estimator reduces to the one proposed for missing data in \cite{ferretti2012neutrality}, which is equivalent to
\beq
P_c(j|c=
n,\{r\})=I\left(j=\sum_{i=1}^{n}I(r_i>0)\right) \ .
\eeq
This choice was implicitly suggested also in \cite{roychoudhury2010sufficiency}.

\subsubsection{Multiple diploid individuals}

In this case all individuals have $c_i=2$ and the MCLE was already derived by Hellmann et al. \cite{hellmann2008population}:
\beq
P_c(j|c=2n,\{r\})=\sum_{i_1=0}^{2}\ldots\sum_{i_n=0}^{2}I\left(j=\sum_{p=1}^n i_p\right)\prod_{q=1}^n P^*(i_q|c_q=2,r_q)\ .
\label{hellmanneq}
\eeq

\subsubsection{Pools}

In this case, there is a single unit of $c$ chromosomes. The probability was derived by Ferretti et al. \cite{ferretti2013}
\beq
P_c(j|c,r)=P^*(j|c,r)
\eeq
but see also Section \ref{sec_dual} for a simpler formula. 

For multiple pools, the probability follows closely equation (\ref{indlineages}) where $c_1\ldots c_U$ are the numbers of (haploid) individuals inside each pool and $r_1\ldots r_U$ are the read depths of the pools at the position considered.

\subsection{Related lineages}

Sequencing unrelated individuals (either pooled together or sequenced separately) is the most common experimental setup for variability studies as described in the previous section, but it is not the only one. 
There are several cases where lineages in different units are related by identity (for example, the same individual sequenced both alone and in a pool with other individuals) or identity-by-descent, like for trios or inbred lines from a population. In this section we develop estimators for these cases. 

\subsubsection{Trios}
A trio is a (diploid) family of mother, father and child that are sequenced separately. We assume that father and mother are two unrelated individuals from the same population. We restrict our analysis to autosomes, where the two alleles of the child are the copies of one paternal and one maternal allele. 
For complete sequences, the probability is just $P(j)=\delta_{j,4}$ since there are four independent lineages. 

For NGS data, we denote by $r_{M}$, $r_{F}$ and $r_{C}$ the read depths of mother, father and child respectively. We obtain $P_c(j|r_M,r_F,r_C)$ by conditioning on the number of lineages $j'$ sequenced in the parents. We can rewrite it in terms of the probability $P_c(j'|c=4,r_M,r_F)$ of the parents alone (eq.(\ref{hellmanneq})):
\beq
P_c(j|r_M,r_F,r_S)=\sum_{j'=0}^4 p_t(j|j',r_M,r_F,r_C)P_c(j'|c=4,r_M,r_F) \label{ptrios}
\eeq
where $p_t(j|j',r_M,r_F,r_C)$ is the probability of sequencing $j$ independent lineages in the trio given the number of independent lineages $j'$ sequenced from the parents. $p_t(j|j',r_M,r_F,r_C)$ is obtained case by case depending on the probability that the sequences of the child could contain new alleles with respect to the parental sequences and the probability to detect them. 
For example, consider the case $j'=3$. Then there is only a single allele in the parents that has not been sequenced. This allele is absent in the child with 50\% probability (in this case $j=3$ because no new alleles are sequenced in the child) or it could be present with 50\% probability, but not sequenced (then $j=3$ with probability $2^{-r_c}$) or could be sequenced (then $j=4$ with probability $1-2^{-r_c}$).

The complete probability is 
\begin{align}
p_t(j|j'=0,r_M,r_F,r_C)=&P^*(j|2,r_S) \\ 
p_t(j|j'=1,r_M,r_F,r_C)=&\frac{1}{2}P^*(j-1|2,r_S)+\frac{1}{2}(\delta_{j,1}2^{-r_S}+\delta_{j,2}(1-2^{-r_S})) \nonumber \\ 
p_t(j|j'=2,r_M,r_F,r_C)=&\frac{1}{2}(1+I(r_Mr_F=0))(\delta_{j,2}2^{-r_S}+\delta_{j,3}(1-2^{-r_S})) + \nonumber\\+ & \frac{1}{4}I(r_Mr_F>0)\left( \delta_{j,2}+ P^*(j-2|2,r_S)\right)\nonumber \\
p_t(j|j'=3,r_M,r_F,r_C)=& \frac{1}{2}\delta_{j,3}+\frac{1}{2}(\delta_{j,3}2^{-r_S}+\delta_{j,4}(1-2^{-r_S}))\nonumber \\ 
p_t(j|j'=4,r_M,r_F,r_C)=& \delta_{j,4}\ .\nonumber
\end{align}

Multiple unrelated trios can be dealt with by replacing the probability $P^*(i_q|\ldots)$ in equation (\ref{indlineages}) with the probability (\ref{ptrios}) and replacing $c_q$ with 4.

\subsubsection{Pooled trios}
A pooled trio is a family of mother, father and child that are pooled together and sequenced. We consider $n$ unrelated families, each family sequenced separately from the others, and denote by $r_i$ the total read depths. $P_c(j|\{r_i\}_{i=1\ldots n})$ is given by
\beq
P_c(j|\{r_i\}_{i=1\ldots n})=\sum_{i_1=0}^{4}\ldots\sum_{i_{n}=0}^{4}I\left(j=\sum_{p=1}^{n} i_p\right)\prod_{q=1}^{n} 
P_{pt}(i_q|r_q)
\eeq
where $P_{pt}(i|r)$ is the probability of sequencing $i$ homolog chromosomes for a single pooled trio. 
It can be derived case-by-case conditioning on the number $i'$ of sequenced chromosomes (identical or not) for a pool,  obtaining
\beq
P_{pt}(i|r)=\sum_{i'}p_{pt}(i|i')P^*(i'|c=6,r)\ .
\eeq
$p_{pt}(i|i')$ can be found by conditioning on the number of non-inherited chromosomes sequenced, obtaining 
\begin{align}
p_{pt}(i|i')&= \delta_{i,i'}\frac{ 4{2\choose i'-2}+4{2\choose i'-1}+{2\choose i'}}{{6\choose i'}}+\delta_{i,i'-1}\frac{ 4{2\choose i'-3}+2{2\choose i'-2}}{{6\choose i'}}+\delta_{i,i'-2}\frac{ {2\choose i'-4}}{{6\choose i'}}
 \end{align}
and finally
\begin{align}
P_{pt}(i|r)&= \left[ 4{2\choose i-2}+4{2\choose i-1}+{2\choose i}\right]\frac{P^*(i|6,r)}{{6\choose i}}+ 
\\ \nonumber 
&+\left[ 4{2\choose i-2}+2{2\choose i-1}\right]\frac{P^*(i+1|6,r)}{{6\choose i+1}}+ {2\choose i-2}\frac{P^*(i+2|6,r)}{{6\choose i+2}}\ .
 \end{align}


\subsubsection{Pools and complete sequences with overlapping individuals}
A potentially useful setup is the combination of complete sequences of few individuals and a pool of several individuals from the same population. In this situation, there could be individuals in the pool for which the complete sequence is also available. 

Here we deal with the haploid case, but the results can be easily adapted to the diploid case by considering a diploid individual as a pair of haploids. 
Denote by $m$ the number of individuals completely sequenced, by $n$ the number of individuals pooled, and by $o$ the overlapping between the two groups of individuals, i.e. the individuals in the pool that have also been sequenced separately. Denote by $r$  the read depth of the pool. 
The distribution of $j$ can be obtained by conditioning on the number $l$ of pooled reads that come  actually from the $n-o$ individuals exclusive to the pool. The distribution of $l$ is a binomial with probability $o/n$ and $r$ extractions, therefore
\begin{align}
P_c(j|r)=&\sum_{l=0}^r  P^*(j-m|n-o,l) { r \choose l } \left(1-\frac{o}{n}\right)^l \left(\frac{o}{n}\right)^{r-l} =\label{poolind} \\
=& \frac{(n-o)!}{n^r(n-o-j+m)!} \sum_{l=0}^r { r \choose l } o^{r-l} S(l,j-m) \ .\nonumber
\end{align}
See also Section \ref{sec_dual} for a simpler formula.


\subsubsection{Inbred lines}
Consider a population from which $n$ inbred lines are derived. We denote the initial heterozygosity by $H$ and the final heterozygosity by $H_{inbred}$. The degree of inbreeding is measured by the inbreeding coefficient $F=(H-H_{inbred})/H$, that is the relative decrease in heterozygosity $H$ due to inbreeding, 
and is assumed to be known. $F$ is also equal to the probability of identity by descent for the inbred line. 
For each line, a diploid individual is sequenced. 
Our aim is to estimate the heterozygosity of the initial population from the sequences of 
individuals from the inbred lines.

If complete sequences are available, since each individual has an independent probability $F$ of being homozygote because of inbreeding, the distribution of the number of homozygotes is just a binomial. But the number of sequenced chromosomes is $2n$ minus the number of homozygotes, therefore 
\begin{align}
&P_c(j|c=2n)={n \choose 2n-j}F^{2n-j}(1-F)^{j-n}\ .\label{inbred1}
\end{align}
If instead we have NGS reads, the distribution of sequenced chromosomes should account for the ``
effective homozygote probability'' $F+(1-F)2^{-r_q+1}$ due to sampling:
\begin{align}
&P_c(j|c=2n,\{r\})=\sum_{i_1=0}^{2}\ldots\sum_{i_n=0}^{2}I\left(j=\sum_{p=1}^n i_p\right)
 \cdot  \prod_{q=1}^n \left[I(r_q=0)\delta_{i_q,0}+ \right. \label{inbred2}\\ &
\left. +I(r_qi_q>0)\left(F+(1-F)2^{-r_q+1}+(i_q-1)\left(2(1-F)(1-2^{-r_q+1})-1\right)\right)\right]\ .\nonumber
\end{align}

Note that in this case, the formula (\ref{watt}) with (\ref{inbred1}), (\ref{inbred2}) can be inverted to give the expected variability $\ev(S)$ for individuals from inbred lines with a given inbreeding coefficient $F$. 

\section{An equivalent form for Watterson estimators}\label{sec_dual}

\subsection{Equivalence between the estimator of Jiang \textit{et al.} and the Watterson estimator for pools}
An unbiased estimator of $\theta$ based on $S$  was proposed in \cite{jiang} for NGS data of multiple diploid individuals, even if this is not the most appropriate setup,  as we will see immediately. The estimator is
\beq
\hat{\theta}_{J}=\frac{S}{\sum_{r=2}^{\infty}{L_r}\sum_{k=1}^{c-1} \frac{1}{k} \left(1-\left(\frac{k}{c}\right)^{r}-\left(1-\frac{k}{c}\right)^{r}\right)} \label{thetaj}
\eeq
where $c$ is twice the sample size (for diploids) and $r$ is the total read depth. This estimator is unbiased since the mean of $S$ is given by the probability $\theta/k$ of a SNP of frequency $k$ in the sample multiplied by the probability of detecting it in a random extraction of $r$ alleles, that is $1-\left(\frac{k}{c}\right)^{r}-\left(1-\frac{k}{c}\right)^r$.

A first observation is that this estimator is not actually unbiased for reads coming from multiple individuals sequenced separately. In fact, it takes into account only the total number of reads, while an unbiased estimator would depend on how they are distributed among individuals.
However, it is an unbiased estimator of $\theta$ for pooled sequences, since in that case information about the origin of the reads is lost. 

Furthermore, there is only a single unbiased estimator proportional to $S$, since the proportionality constant is fixed by the bias of $S$. This means that the estimator $\hat{\theta}_{J}$ is actually the Watterson estimator $\hat{\theta}_W$ for pools proposed in \cite{ferretti2013}. 
The two different forms derive from different intermediate conditioning for $Z$: on the allele frequency $k$ in the sample in the first case, on the number of lineages actually sequenced $j$ in the second.

Note that in the light of this equivalence, the conclusions of \cite{jiang} about the differences between their estimator and Hellmann's one when applied to individual data are at least doubtful. 
They found both estimators to be biased and the variance of Hellmann's one to be significantly larger, but theory suggests that they are unbiased and the variance of Hellmann's one should be lower. In fact,  numerical simulations performed in \cite{ferretti2013} showed almost no bias, no sensible difference in variance and a very good correlation between them.

From the mathematical point of view, the equality between $\hat{\theta}_{J}$ and $\hat{\theta}_{W}$ for pools and the related equalities that we will present in the next section depend on a 
family of combinatorial identities. 
The identity of the two estimators $\hat{\theta}_J$ and $\hat{\theta}_W$ for pools implies identity of their denominators. The reasoning in this section 
is equivalent to a double counting proof of the combinatorial identity
\beq
\sum_{j=1}^{\min(c,r)}\frac{c!}{(c-j)!}S(r,j)a_j=\sum_{k=1}^{c-1}\frac{c^{r}-k^{r}-\left(c-k\right)^{r}}{k} \label{combid}
\eeq
valid for all pairs of integers 
$(c,r)$ such that $c\geq 1$ and $r\geq 1$. 
The identity involves Stirling numbers $S(r,j)$ and harmonic numbers $a_j$ in a nontrivial way. Note that both sides of the identity are integers. This identity can also be proved directly [M.Mamino, persona communication].
This identity is a combination of a family of related identities for a general spectrum, presented in \ref{appcomb}.


\subsection{General alternative form for the Watterson estimators}
The above form of \cite{jiang} for the Watterson estimator for pools can be generalized to the whole family of estimators for units of independent lineages, described by equations (\ref{thetaw}) and (\ref{indlineages}). We follow the same notation as before, but we denote the total number of lineages by $c=\sum_{i=1}^U c_i$. The general form for these estimators is
\begin{align}
&\hat{\theta}_{W}=\frac{S}{\sum_{\{r\}}L_{\{r\}} \sum_{k=1}^{c-1} \frac{1}{k} \Pi_k(\{c\},\{r\}) } \ ,\label{wattalt} \\
&\Pi_k(\{c\},\{r\})= \sum_{k_1=0}^{c_1}\ldots \sum_{k_U=0}^{c_U} I\left(k=\sum_{i=1}^U k_i\right) \frac{\prod_{i=1}^U {c_i \choose k_i}}{ {c \choose k} } 
\left[1-\prod_{i=1}^U \left( 
\frac{k_i}{c_i} \right)^{r_i} -\prod_{i=1}^U \left(1-\frac{k_i}{c_i} \right)^{r_i}
\right]\nonumber
\end{align}
where the multi-hypergeometric distribution  ${\prod_{i=1}^U {c_i \choose k_i}}/{ {c \choose k} }$ describes how the alleles are assigned to the different units and the term $\prod_{i=1}^U \left( 
\frac{k_i}{c_i} \right)^{r_i} + \prod_{i=1}^U\left(1-\frac{k_i}{c_i} \right)^{r_i}
$ is the probability of extracting just one of the two alleles. 
All the estimators of section \ref{probindlin} can be rewritten in this form. This form is often more convenient computationally than the combinatorics in equations (\ref{thetaw}), (\ref{indlineages}).

For a generic frequency spectrum $\ev(\xi_k|n)=\theta L \bar{\xi}_{k,n}$,  equation (\ref{wattalt}) should be replaced by 
\beq
\hat{\theta}_{W}=\frac{S}{\sum_{\{r\}}L_{\{r\}} \sum_{k=1}^{c-1} \bar{\xi}_{k,n} \Pi_k(\{c\},\{r\}) } \ .\label{wattaltgen}
\eeq


We can also find an estimator  similar to (\ref{thetaj}) for a combination of a pool of $n$ (haploid) individuals and $m$ complete sequences, $o$ of which are overlapping. In this case simple combinatorial reasoning on the probability of detecting a SNP of frequency $k$ among the $n+m-o$ individuals (the SNP is detected unless all complete sequences share the same allele) leads to the unbiased estimator
\beq
\hat{\theta}_{W}=\frac{S}{\sum_{r=2}^{\infty}{L_r}\sum_{k=1}^{n+m-o-1}  \left(1-\frac{{ n-o \choose k}}{{ n+m-o\choose k}}\left(1-\frac{k}{n}\right)^{r}-\frac{{ n-o \choose k-m }}{{ n+m-o\choose k}}\left(\frac{k-m+o}{n}\right)^{r}\right)\frac{1}{k}} 
\eeq
that is equivalent to the case (\ref{poolind}) of estimator (\ref{thetaw}). 

\section{Watterson estimators for autopolyploids}\label{sec_poly}

A particularly interesting and challenging set of data is represented by polyploid genomes. Species with ploidy greater than 2 are highly interesting from an evolutionary point of view, as well as economically in agrobiotech and breeding since it involves many commercial species of plants (e.g. potato, sugar cane) and fishes (e.g. Salmonidae).

Polyploid species are difficult both to sequence and to analyze, due to the complex homology/paralogy relation between the constituent genomes. However, some polyploids can be treated by the methods developed here. In particular, autopolyploids are polyploid organisms with different homologous chromosomes from the same species. 
Autotetra
ploid populations follow the standard coalescent as shown by \cite{Arnoldetal2012}, 
and this can be extended to autopolyploids that have similar transition probability matrices
. Here we present estimators of variability for populations of autopolyploid species.

Polyploids can be considered as pools with number of lineages equal to their ploidy. Multiple polyploids can then be considered as combinations of pools, but they can be pooled themselves. We consider a species with ploidy $p$.
The estimators for autopolyploids are given by 
\beq
\hat{\theta}_{W}=\frac{S}{\sum_{r=2}^{\infty}{L_r}\sum_{k=1}^{p-1} \frac{1}{k} \left(1-\left(\frac{k}{p}\right)^{r}-\left(1-\frac{k}{p}\right)^{r}\right)} \label{thetapoly}
\eeq
for a single polyploid individual, where $r$ is the read depth, and by the formula
\begin{align}
&\hat{\theta}_{W}=\frac{S}{\sum_{\{r\}}L_{\{r\}} \sum_{k=1}^{np-1} \frac{1}{k} \Pi_k(n,p,\{r\}) } \ ,\label{thetamultipoly} \\
&\Pi_k(n,p,\{r\})= \sum_{k_1=0}^{p}\ldots \sum_{k_n=0}^{p} I\left(k=\sum_{i=1}^n k_i\right) \frac{\prod_{i=1}^n {p \choose k_i}}{ {np \choose k} } \left[1-\prod_{i=1}^n 
\left(\frac{k_i}{p} \right)^{r_i} -\prod_{i=1}^n \left(1-\frac{k_i}{p} \right)^{r_i} 
\right]\nonumber
\end{align}
for sequences from $n$ polyploid individuals, where $\{r\}=\{r_i\}_{i=1\ldots n}$ are the read depths per individual. This is also equivalent to the formula (\ref{thetaw}) for $\hat{\theta}_W$ with
\beq 
P_c(j|\{r\})=\sum_{i_1=0}^{p}\ldots\sum_{i_{n}=0}^{p}I\left(j=\sum_{l=1}^{n} i_l\right)\prod_{q=1}^{n} P^*(i_q|p,r_q)\ .
\eeq

The estimator for a pool of polyploid individuals is the same as in the general case for pools with $c=np$, where $n$ is the number of individuals in the pooled sample:
\beq
\hat{\theta}_{W}=\frac{S}{\sum_{r=2}^{\infty}{L_r}\sum_{k=1}^{np-1} \frac{1}{k} \left(1-\left(\frac{k}{np}\right)^{r}-\left(1-\frac{k}{np}\right)^{r}\right)}\ . \label{thetapolypool}
\eeq


\section{Discussion}

In this paper we have presented a large family of generalized Watterson estimators that are suited for different types of NGS data, from haploids to polyploids, pools and trios, or a mix of NGS/Sanger data. These estimators are built on the Maximum Composite Likelihood approach; furthermore they are unbiased and depend linearly on $S$, which is a sufficient statistic for small $\theta$. 
The general theory presented here includes all these estimators and many others. Existing estimators are assigned to the proper place in this unified framework. 

We pay special attention to estimators for single and multiple autopolyploid individuals. Sequencing of these species has proved to be hard, but more and more projects will soon be devoted to some of the more interesting polyploid species from a commercial point of view, especially among domesticated plants \cite{Brenchley2012,HanKang2011,WangWang2011}.
Autopolyploids without a strong inbreeding follow the dynamics of the usual coalescent, so our theory is applicable to these species. 
On the other hand, allopolyploids (whose genome derives from different species) cannot be studied by the same technique since the differences between homologous chromosomes from different constituent species are much stronger and the divergence time between them is often of order of the divergence between species. 
Specific methods have to be developed for the analysis of variability in allopolyploids \cite{Justin2013,Brenchley2012}. 
A simple approach could be the study of the variability of each constituent genome and, independently, the genetic differentiation between them.

We did not discuss an important issue with NGS data, that is, base errors. Sequencing errors and misalignments occur at an high rate in NGS data. The bases with lower quality can be removed from the reads or the sequences, however sequencing errors or similar effects can often generate false SNPs at low frequency and it could be difficult to distinguish them from true low frequency alleles. 
In this case, filtering or SNP calling is usually applied to the data, resulting in an unknown $Z_{\varphi,\xi}$ for these alleles. Denote by $(\Phi,\Xi)_\epsilon$ the set of features $\{\varphi,\xi\}$ strongly affected by sequencing errors. In some cases it is possible to estimate $Z_{\varphi,\xi}$ or to correct $S_{\varphi,\xi}$ based on quality scores for called SNPs or known error rates. On the other hand, if it is not possible to estimate the contribution of the errors, a good practice is to discard the corresponding $S_{\varphi,\xi}$, $\{\varphi,\xi\}\in(\Phi,\Xi)_\epsilon$ and to work with the approximate MCL estimator for this case, that is
\beq
\hat{\theta}_W=\frac{\sum_{\{\varphi,\xi\}\notin(\Phi,\Xi)_\epsilon}S_{\varphi,\xi}}{\sum_{\{\varphi,\xi\}\notin(\Phi,\Xi)_\epsilon}L_\varphi Z_{\varphi,\xi}}\label{eq_rem}
\eeq
as proposed in \cite{achaz2008testing} for sequence data and \cite{Futschik:2010fk},\cite{ferretti2013} for pooled reads. The only alternative is to estimate $Z_{\varphi,\xi}$ by heuristic methods.

Generalizing equation (\ref{eq_rem}), it is also possible to extend the results of this paper to generic sums of the frequency spectrum, for example estimators of the form $\sum_{i=1}^k\xi_k/\sum_{i=1}^k1/k$ which consider only the lowest frequencies. 

The estimator proposed here assume the standard Wright-Fisher neutral model for the allele frequency spectrum. 
However, an arbitrary expected frequency spectrum $\ev({\xi}_k|n)=\theta L\bar{\xi}_{k,n}$ could be used in the place of the neutral spectrum $\theta L/k$. 
It is sufficient to replace $a_j$ by $\sum_{i=1}^{j-1}\bar{\xi}_{i,j}$ in the denominator of equation (\ref{thetaw}) or to replace $1/k$ by $\bar{\xi}_{k,c}$ in the denominator of equation (\ref{wattalt}). 
This extends previous adaptations of the original Watterson estimator to null scenarios with demography or varying population size (e.g. \cite{ZivkovicWiehe}, \cite{ferretti2010optimal}).

In this paper we used the composite likelihood approximation to derive the estimators of variability. However, the variance of these Watterson estimators depends on recombination. 
The usual formulae for ML work only for unlinked sites. In this case, in the limit $\theta\rightarrow 0$ and $\theta L$ constant, $S_{\varphi,\xi}$ is Poisson distributed, i.e.  $\var(S_{\varphi,\xi})=\ev(S_{\varphi,\xi})$ and therefore $\var(\hat{\theta}_W)=\theta/\sum_{\varphi\in\Phi}L_\varphi Z_\varphi$. 
In the same limit, the variance for linked sites contain a term $\theta^2L^2$ coming from the covariances between sites \cite{fu1995statistical,ferretti2012neutrality,ferretti2013}. An exact formula for this term of the variance is available only for a few cases: complete sequences, sequences with missing data \cite{ferretti2012neutrality} and pooled NGS reads \cite{ferretti2013}. 
In the case of completely linked sites and known variance, these estimators could be improved \cite{fu1994estimating}, also by shrinkage methods \cite{futschik2008inadmissibility}. Note that the variance of the MCLE could be estimated by the bootstrapping methods described in \cite{roy}.

Finally, the theoretical framework developed in this paper allowed to obtain an interesting set of combinatorial identities. This is another example of the way research on theoretical population genetics is highly connected to some fields of mathematics, e.g. combinatorics \cite{arratia2003logarithmic} and could lead to further mathematical insights. 

\section*{Acknowledgments}
We thank A. Fonseca Amaral, M. P\'erez-Enciso, W. Burgos, B. Nevado and G. Achaz for useful discussions, M. Mamino for providing an explicit proof of the main combinatorial identity and two anonymous referees for their constructive comments. LF acknowledges support from ANR-12-JSV7-0007 (ANR, France). The project was funded by Grants CGL2009-09346 (MICINN, Spain) and AGL2013-41834-R (MEC, Spain) to SERO and by a Consolider Grant
from Spanish Ministry of Research, CSD2007-00036 ``Centre for Research in Agrigenomics''.

\appendix 
\section{Combinatorial identities}\label{appcomb}

%

We can extend the previous identity (\ref{combid}) to a set of identities derived from the same equivalence of estimators but for an arbitrary frequency spectrum. The fundamental identities are obtained by double counting technique. 

We considering a frequency spectrum concentrated around a single frequency $\tilde{f}$ in the population (i.e. $\xi(f)=\delta\left(f-\tilde{f}\right)$, or $\xi_k={n \choose k} \tilde{f}^k\left(1-\tilde{f}\right)^{n-k}$ for the sample spectrum). By double counting, the two Watterson estimators of the form (\ref{thetawgen}) and (\ref{wattaltgen}) should be equal, and therefore we can equal their denominators. By computing the Taylor expansion in the variable $\tilde{f}$ of both sides and equating the coefficients of the $l$th power, we obtain:
\begin{align}
\sum_{j=2}^{\min(c,r)}\frac{c!}{(c-j)!}S(r,j)&\sum_{k=1}^{j-1} (-1)^k{j \choose k, l-k, j-l} =\nonumber \\  
&\sum_{k=1}^{c-1} (-1)^k{c \choose k,l-k,c-l} \left({c^{r}-k^{r}-\left(c-k\right)^{r}}\right) \label{combid2}
 \end{align}
for integers $(r,c,l)$ with $r\geq 1$ and $1\leq l\leq c$. They involve Stirling numbers and multinomials. Any other identity in this family (including (\ref{combid})) can be obtained as a linear combinations of these ones. Note that since the l.h.s. is 0 for $l>r$, these identities reduce to 
\beq
\sum_{k=1}^{l} (-1)^k{c \choose k,l-k,c-l} \left({c^{r}-k^{r}-\left(c-k\right)^{r}}\right) =0
\eeq
for $r<l\leq c$.

%




\section*{References}

\bibliographystyle{elsarticle-num}
\bibliography{popgenwatt}







\end{document}